\documentclass[12pt]{article}

\advance \textheight by \topskip
\usepackage{amsmath,amssymb}
\begin{document}
\title{
\begin{flushright}
{\small USACH-FM-01/10}\\[1.0cm]
\end{flushright}
{\bf Nonlinear Supersymmetry\footnote{Plenary talk given by
M. P. at the Fourth Inter.
Conference
"Symmetry in Nonlinear Mathematical Physics", July 9-15,
2001, Kiev, Ukraine
(Published in Proceedings,
Editors A. G. Nikitin, V. M. Boyko and R. O. Popovych, Kyiv,
Inst. of Mathematics, 2002, V.43, Part 2, pp. 508--519)
}
}}

\author{{\sf Sergey M. Klishevich${}^{a,b}$}\thanks{
E-mail: sklishev@lauca.usach.cl}
{\sf\ and Mikhail S. Plyushchay${}^{a,b}$}\thanks{
E-mail: mplyushc@lauca.usach.cl}
\\
{\small {\it ${}^a$Departamento de F\'{\i}sica,
Universidad de Santiago de Chile,
Casilla 307, Santiago 2, Chile}}\\
{\small {\it ${}^b$Institute for High Energy Physics,
Protvino, Russia}}}
\date{}

\maketitle

\vskip-1.0cm

\begin{abstract}
 After a  short   discussion of the intimate relation
 between the generalized statistics and supersymmetry, we
 review the recent  results  on the nonlinear supersymmetry
 obtained in the context of the quantum anomaly problem  and
 of the universal algebraic construction associated with the
 holomorphic nonlinear supersymmetry.
\end{abstract}
\newpage

\section*{Introduction}
Nonlinear supersymmetry is a natural generalization of the
usual linear supersymmetry
\cite{plyushchay:witten,plyushchay:cooper&khare&sukhatme}.
It is realized variously in such different systems  as
the parabosonic
\cite{plyushchay:plyushchay00} and
parafermionic
\cite{plyushchay:klishevich&plyushchay99}
oscillator models, and the $P,T$-invariant models
of planar fermions
\cite{plyushchay:grignani&plyushchay&sodano} and
Chern-Simons fields
\cite{plyushchay:nirov&plyushchay98}. It
is also the symmetry  of
the fermion-monopole system
\cite{plyushchay:jonghe&macfarlane&peeters,
plyushchay:plyushchay00a}.
The algebraic structure of the nonlinear supersymmetry
resembles the structure of the finite $W$-algebras
\cite{plyushchay:boer&harmsze&tjin}
for which the commutator of
generating elements is proportional to a finite
order polynomial in them. In the simplest case the
nonlinear supersymmetry is characterized by the
superalgebra of the form
\begin{align}\label{plyushchay:NSUSY}
 [Q^\pm,H]&=0,& (Q^\pm)^2&=0,& \{Q^+,Q^-\}=P_n(H),&&
\end{align}
where $P_n(.)$ is a polynomial of the $n$-th degree. The
nonlinear supersymmetry with such a superalgebra was
investigated for the first time by Andrianov, Ioffe and
Spiridonov
\cite{plyushchay:andrianov&ioffe&spiridonov}.

The pseudoclassical construction
underlies the supersymmetric quantum mechanics of Witten
\cite{plyushchay:witten,plyushchay:cooper&khare&sukhatme}
corresponding to the linear ($n=1$) case of  the
superalgebra (\ref{plyushchay:NSUSY}).
Though the nonlinear supersymmetry can also be realized
classically, there is an essential difference from  the
linear case:
the attempt to quantize the nonlinear supersymmetry
immediately faces the problem of the quantum anomaly
\cite{plyushchay:plyushchay00,
plyushchay:klishevich&plyushchay01a}.
It was shown
\cite{plyushchay:klishevich&plyushchay01b} that
the universal algebraic structure with
associated ``integrability conditions'' in the form of
the Dolan-Grady relations
\cite{plyushchay:dolan&grady}
underlies the so called holomorphic nonlinear supersymmetry
\cite{plyushchay:klishevich&plyushchay01a}.
This  structure allows ones to find  a
broad class of anomaly-free quantum mechanical systems
related to the exactly and quasi-exactly solvable systems
\cite{plyushchay:ulyanov&zaslavskii,
plyushchay:turbiner88,plyushchay:shifman,
plyushchay:ushveridze,
plyushchay:finkel&gonzalez&kamran&olver&rodriguez,
plyushchay:bender&dunne},
and gives a  nontrivial centrally extended
generalization of the superalgebra (\ref{plyushchay:NSUSY})
\cite{plyushchay:klishevich&plyushchay01b}.

In this  talk, after a  short   discussion of the intimate
relation  between the generalized statistics and
supersymmetry
\cite{plyushchay:plyushchay00}, we shall
review the recent
results  on the nonlinear supersymmetry obtained in the
context of the quantum anomaly problem  and of the
universal algebraic construction associated with  the
holomorphic nonlinear supersymmetry
\cite{plyushchay:klishevich&plyushchay01a,
plyushchay:klishevich&plyushchay01b}.

\section*{Nonlinear supersymmetry in purely parabosonic
 \protect\\systems}
Some time ago it was shown that the linear supersymmetry
can be realized without fermions
\cite{plyushchay:plyushchay96,plyushchay:plyushchay96a,
plyushchay:gamboa&plyushchay&zanelli}.
The nonlinear supersymmetry admits a similar realization
revealing the close relationship between the generalized
statistics and supersymmetry
\cite{plyushchay:plyushchay00}.
The  relationship can be observed in the following way.
Let us consider a  single-mode paraboson system defined by
the relations
\begin{align*}
 [\{a^+,a^-\},a^\pm]&=\pm a^\pm,&
 a^-a^+|0\rangle&=p|0\rangle,& a^-|0\rangle&=0,&&
\end{align*}
where
$p\in\mathbb N$ is the order of a paraboson
\cite{plyushchay:ohnuki&kamefuchi}.
Then the direct calculation shows that the
pure parabosonic system of the even order
$p=2(k+1)$, $k\in\mathbb Z_+$,
with the  Hamiltonian of the simplest quadratic form
$H=a^+a^-$
reveals a spectrum typical  for the  nonlinear
supersymmetry:
all its  states are paired in doublets except
the $k+1$ singlet states
$|2l\rangle\propto(a^+)^{2l}|0\rangle$,
$l=0,\ldots,k$.
In correspondence with this property,
the system has two integrals of motion
\begin{align}\label{plyushchay:nonlocQ}
 Q^+&=(a^+)^{2k+1}\sin^2\frac\pi 4\{a^+,a^-\},&
 Q^-&=(a^-)^{2k+1}\cos^2\frac\pi 4\{a^+,a^-\},
\end{align}
which together with the Hamiltonian form
the nonlinear superalgebra
(\ref{plyushchay:NSUSY}) of the order $n=2k+1$ with
$P_{2k+1}(H)=H\cdot \prod_{m=1}^k(H^2-4m^2)$.
This simplest system reflects the peculiar
feature of the parabosonic realization of supersymmetry:
the supercharges  are realized  in the form
of the infinite series in $a^\pm$, and
the role of the grading operator is played here
by $R=(-1)^N=\cos \pi N$, where
$N=\frac 12\{a^+,a^-\}-\frac 12p$ is the
parabosonic number operator.

It is known that the  deformed Heisenberg algebra with
reflection
\begin{align}
\label{plyushchay:RDHA}
 [a^-,a^+]&=1+\nu R,&\{R,a^\pm\}&=0,&R^2&=1,&&
\end{align}
underlies the parabosons
\cite{plyushchay:macfarlane,plyushchay:plyushchay97}.
This algebra
possesses unitary infinite-dimensional
representations for $\nu>-1$, and at the integer values of
the deformation parameter, $\nu=p-1$, $p\in\mathbb N$, is
directly related to parabosons of order $p$
\cite{plyushchay:ohnuki&kamefuchi,plyushchay:macfarlane,
plyushchay:plyushchay97}. On the other hand, at $\nu=-(2p+1)
$
the $R$-deformed Heisenberg algebra has finite-dimensional
representations corresponding to the deformed parafermions
of order $2p$
\cite{plyushchay:plyushchay97}.
In the coordinate representation the operator $R$
is the parity operator and
the operators $a^\pm$ can be realized in the form
$a^\pm=\frac 1{\sqrt 2}(x\mp iD_\nu)$ with
$D_\nu=-i(\frac d{dx}-\frac\nu{2x}R)$.
In the context of the Calogero-like models,
the operator $D_\nu$ is known as the Yang-Dunkl
operator
\cite{plyushchay:yang,plyushchay:dunkl},
where $R$ is treated as  the
exchange operator. In the coordinate  representation the
Hamiltonian $H=a^+a^-$
and supercharges (\ref{plyushchay:nonlocQ}) read as
\cite{plyushchay:plyushchay00}
\begin{align}
 H&=\frac 12\left(-\frac{d^2}{dx^2}+x^2+\frac{\nu^2}{4x^2}
 -1+\nu\left(\frac 1{2x^2}-1\right)R\right),
 \label{plyushchay:c1} \\
 Q^+&=(Q^-)^{\dag}=
 \frac 1{2^{3(k+\frac 12)}}\left(\left(-\frac{d}{dx}+x
 +\frac{\nu}{2x}\right)(1-R)\right)^{2k+1}
 \label{plyushchay:c1q}
\end{align}
with $\nu=2k+1$. The system given by the Hamiltonian
(\ref{plyushchay:c1}) can be treated as a 2-particle
Calogero-like
model with exchange interaction, where $x$ has a sense
of a relative coordinate and $R$ has to be understood as the
exchange operator
\cite{plyushchay:polychronakos92,
plyushchay:polychronakos93}.
Therefore, at odd values of the parameter $\nu$, the class
of Calogero-like systems (\ref{plyushchay:c1}) possesses a
hidden
supersymmetry, which at $\nu=1$ is the linear ($n=1$)
supersymmetry in the unbroken phase, whereas at $\nu=2k+1$,
the supersymmetry is characterized by the supercharges being
differential operators of order $2k+1$ satisfying the
nonlinear superalgebra~(\ref{plyushchay:NSUSY}).
Recently the realization of the nonlinear supersymmetry
was extended within the standard approach with fermion
degrees of freedom to the case of  multi-particle
Calogero and related  models
\cite{plyushchay:sasaki&takasaki}.

\section*{Classical supersymmetry}

Let us turn now to the classical formulation of the
supersymmetry (\ref{plyushchay:NSUSY}).
For the purpose, we consider  a
non-relativistic particle in one dimension
described by the Lagrangian
\begin{equation}
\label{plyushchay:Lgen}
 L=\frac 12\dot x^2-V(x)-L(x)N+i\theta^+\dot\theta^-,
\end{equation}
where $\theta^\pm$ are the Grassman variables,
$(\theta^+)^*=\theta^-$, $N=\theta^+\theta^-$, and
$V(x)$ and $L(x)$ are two real functions. The
nontrivial Poisson-Dirac brackets for the system are
$\{x,\,p\}_*=1$ and $\{\theta^+,\,\theta^-\}_*=-i$, and the
Hamiltonian is
\begin{equation}\label{plyushchay:Hgen}
 H=\frac 12p^2+V(x)+L(x)N.
\end{equation}
The Hamiltonian $H$ and the nilpotent quantity $N$ are
the even integrals of motion for any choice of the
functions $V(x)$ and  $L(x)$, and one can put the question:
when the system (\ref{plyushchay:Lgen}) has also local in
time
odd integrals of motion of the  form
$Q^\pm=B^\mp(x,\,p)\theta^\pm$, where $(B^+)^*=B^-$?
It is obvious that such odd integrals can exist only for a
special choice of the functions $V(x)$ and $L(x)$.
Restricting ourselves to the physically interesting class of
the systems given by the potential $V(x)$ bounded from
below, we can generally represent it in terms of a
superpotential $W(x)$: $V(x)=\frac 12W^2(x)+v$,
$v\in\mathbb R$. Then
all the supersymmetric systems
are separated into the three classes
defined by the behaviour of the superpotential
and the results
can be summarized as follows
\cite{plyushchay:klishevich&plyushchay01a}.

\vskip 1mm
\noindent
{\bf i)} When the physical domain given by $z=W(x)+ip$
includes the origin $z=0$ ($a<W(x)<b$, $a<0$, $b>0$),
the corresponding supersymmetric system
is characterized by the Hamiltonian and the supercharges
of the form
\begin{align}\label{plyushchay:h1}
 H&=\frac{p^2}2+\frac 12W^2(x) + v
 +W'(x)\left[n+W(x)M(W^2(x))\right]N,
 \\[2mm]\notag
 Q^+& =(Q^-)^*=
 z^ne^{i\int_0^pM(p^2-y^2+W^2(x))\,dy}\theta^+, \qquad
 n\in\mathbb Z,
\end{align}
where $M(W^2)$ is an arbitrary regular function,
$\vert M(0)\vert<\infty$. The appearance of the integer
parameter illustrates in this case
the known classical ``quantization
phenomenon''
\cite{plyushchay:nirov&plyushchay97}.
The  appropriate canonical transformation
reduces the system with these Hamiltonian
and supercharges
to the form of the supersymmetric system
with the holomorphic supercharges
\cite{plyushchay:klishevich&plyushchay01a}:
\begin{align}\label{plyushchay:hg}
 H&=\frac 12p^2+\frac 12W^2(x)+v
 +nW'(x)\theta^+\theta^-,&
 Q^+&=(Q^-)^*=z^n\theta^+, & n&\in\mathbb Z_+.&&
\end{align}
The integrals (\ref{plyushchay:hg})
obey the classical nonlinear  superalgebra:
\begin{align}\label{plyushchay:aclas}
  \left\{Q^-,\,Q^+\right\}_*&=-iH^n,&\{Q^\pm,H\}_*&=0.&&&&
\end{align}
The presence of the integer number $n$
in the Hamiltonian means that the instant
frequencies of the oscillator-like odd, $\theta^\pm$,
and even, $z$, $\bar z$, variables are commensurable.
Only in this case the regular odd integrals of motion
can be constructed, and the factor $z^n$
in the supercharges $Q^\pm$ corresponds to the $n$-fold
conformal mapping of the complex plane (or the strip
$a<\mathop{\sf Re} z<b$) on itself (or on the
corresponding region in $\mathbb C$).

\vskip 1mm
\noindent
{\bf ii)} The physical domain is defined by the condition
$\mathop{\sf Re}z\geq 0$ (or $\mathop{\sf Re}z\leq 0$) and
also includes the origin of the complex plane. But unlike
the previous case, there are no closed contours around
$z=0$. In this case the most general form of the Hamiltonian
and the supercharge is
\begin{align}\label{plyushchay:h2}
 H&=\frac{p^2}2+\frac 12W^2(x)+v
 +W'(x)\left[\alpha+R(W(x))\right]N,&
 Q^+&=z^\alpha e^{i\int _{\varphi _0}^\varphi
 R(\rho\cos\lambda)\,d\lambda}\theta^+,
\end{align}
where $\alpha\in\mathbb R$, and we assume that
the function $R(W)$ is analytical at $W=0$ and
\mbox{$R(0)=0$}.
By the canonical transformation
\cite{plyushchay:klishevich&plyushchay01a}, the
Hamiltonian and the supercharges can be
reduced to
\begin{align*}
 H&=\frac 12p^2+\frac 12W^2(x)+v+
 \alpha W'(x)\theta^+\theta^-, &
 Q^+&=(Q^-)^*=z^\alpha\theta^+,&\alpha&\in\mathbb R_+.&&
\end{align*}

\noindent
{\bf iii)} The physical domain is defined by the condition
$\mathop{\sf Re}z>0$ (or $\mathop{\rm Re}z<0$),
i.e. the origin of the complex plane is not included.
Though in this case  the Hamiltonian and
the supercharges have a general form
\begin{align}\label{plyushchay:h3}
 H&=\frac{p^2}2+\frac 12W^2+v+W'(x)\phi(W(x))N,&
 Q^+&=(Q^-)^*=f(H)e^{i\int_{\varphi_0}^\varphi
 \phi(\rho\cos\lambda)\,d\lambda}\theta^+,&
\end{align}
where $\phi$ is some function, the appropriate
canonical transformation reduces it to
\cite{plyushchay:klishevich&plyushchay01a}
\begin{equation*}
 H=\frac 12 p^2+\frac 12 W^2(x)+v,
 \qquad
 Q^\pm=\theta^\pm.
\end{equation*}
This means that {\it classically} the supersymmetry of any
system with
bounded non-vanishing superpotential has a ``fictive''
nature.

In what follows we will
refer to  the nonlinear supersymmetry
generated by the holomorphic supercharges
with  the Poisson bracket (anticommutator)
being  proportional to the  $n$-th order polynomial
in  the Hamiltonian as to the
{\it holomorphic $n$-supersymmetry}.

Though  the
form of the Hamiltonians
(\ref{plyushchay:h1}), (\ref{plyushchay:h2}), and
(\ref{plyushchay:h3}) can be
simplified by applying in every case the
appropriate canonical transformation
reducing the associated supercharges to
the holomorphic or antiholomorphic form,
the quantization  breaks
the equivalence between the corresponding classical systems
(even in the linear case $n=1$)
\cite{plyushchay:klishevich&plyushchay01a}.
Moreover, alternative classical forms for the
Hamiltonians and associated supercharges
are important because of the
quantum anomaly problem to be
discussed below.
Having in mind the importance of alternative classical
formulations
of the nonlinear supersymmetry from the viewpoint of
subsequent
quantization, one can look for the
classical formulation characterized by  the supercharges
of the
$n$-th degree polynomial form in $p$
\cite{plyushchay:klishevich&plyushchay01a}.
The problem of finding such a formulation
can be solved completely in
the simplest case $n=2$, for which the
supercharges are given by
\begin{equation}\label{plyushchay:CalogeroQ}
 Q^\pm=\frac 12\left[\left({}\pm ip+W(x)\right)^2
 +\frac c{W^2(x)}\right]\theta^\pm,\qquad c\in\mathbb R,
\end{equation}
while the Hamiltonian is
\begin{equation} \label{plyushchay:CalogeroH}
 H=\frac 12\left[p^2+W^2(x)
 -\frac{c}{W^2(x)}\right]+2W'(x)N+v.
\end{equation}
Note that the Hamiltonian (\ref{plyushchay:CalogeroH}) has
the
Calogero-like form: at $W(x)=x$ its projection to the unit
of Grassman algebra takes the form of the Hamiltonian of
the two-particle Calogero system. Depending on the value of
the parameter $c$, classically the Calogero-like $n=2$
supersymmetric system (\ref{plyushchay:CalogeroH}) is
symplectomorphic
to the  holomorphic $n$-supersymmetry with
$n=0$ ($c>0$), $n=1$ ($c<0$) or $n=2$ ($c=0$)
\cite{plyushchay:klishevich&plyushchay01a}.

\section*{Quantum anomaly and quasi-exactly solvable (QES)
systems}
According to the results on the supersymmetry in
pure parabosonic systems, a priori one can not exclude the
situation characterized by the supercharges to be the
nonlocal operators represented in the form of some infinite
series in the operator $\frac{d}{dx}$. Since such nonlocal
supercharges have to anticommute for some function of the
Hamiltonian being a usual local differential operator of the
second order, they have to possess a very peculiar
structure.
We restrict ourselves by the discussion
of the supersymmetric systems with the supercharges being
the differential operators of order $n$. Classically this
corresponds to the system (\ref{plyushchay:hg}) with the
holomorphic
supercharges or to the system
(\ref{plyushchay:CalogeroH}).

In the simplest case of the superoscillator possessing the
nonlinear $n$-supersymmetry and characterized by the
holomorphic supercharges of the form (\ref{plyushchay:hg})
with $W(x)=x$,
the form of the classical superalgebra
$\{Q^+_n,Q^-_n\}=H^n$ is changed for
$\{Q^+_n,Q^-_n\}=H(H-\hbar)(H-2\hbar)\ldots
(H-\hbar(n-1))$.
Moreover, it was pointed out in
\cite{plyushchay:plyushchay00}
that for $W(x)\neq ax+b$
a global quantum anomaly arises in a generic
case: the direct quantum analogues of the superoscillators
and the Hamiltonian  do not commute,
$[Q^\pm_n,H_n]\neq 0$.
Therefore, we arrive at the problem of looking for the
classes of superpotentials and corresponding
quantization prescriptions leading to the anomaly-free
quantum
$n$-supersymmetric systems.

Let us begin with the quantum supercharges in the
holomorphic form corresponding to the classical
$n$-supersymmetry,
\begin{equation}\label{plyushchay:Qf}
 Q^\pm=(A^\mp)^n\theta^\pm,\qquad
 \text{where}\quad
 A^\pm={}\mp\hbar \frac{d}{dx}+W(x).
\end{equation}
Choosing the quantum Hamiltonian in the form
(\ref{plyushchay:Hgen}), from the requirement of
conservation of the
supercharges, $\left[Q^\pm,\,H\right]=0$, one arrives at the
supersymmetric quantum system given by the Hamiltonian
\cite{plyushchay:klishevich&plyushchay01a}
\begin{align}\label{plyushchay:Hq1}
 &H=\frac 12\left(-\hbar^2\frac{d^2}{dx^2}+W^2(x)+
 2v+n\hbar\sigma_3W'\right),&
 W(x)&=w_2x^2+w_1x +w_0.
\end{align}
For any other form of the superpotential, the nilpotent
operators (\ref{plyushchay:Qf}) are not conserved.
The family of supersymmetric systems
(\ref{plyushchay:Hq1}) is reduced to the superoscillator at
$w_2=0$
with the associated exact $n$-supersymmetry
\cite{plyushchay:plyushchay00}.
For $w_2\neq 0$, the $n$-supersymmetry is realized always in
the spontaneously broken phase since in this case the
supercharges (\ref{plyushchay:Qf}) have no zero modes
(normalized eigenfunctions of zero eigenvalue).

One can also look for the supercharges in the form of
polynomial of order $n$ in the oscillator-like operators
$A^\pm$ defined in (\ref{plyushchay:Qf}):
\begin{equation}\label{plyushchay:QP}
 Q^\pm=(A^\mp)^n\theta^\pm+
 \sum_{k=0}^{n-1} q_{n-k}(A^\mp)^k\theta^\pm,
\end{equation}
where $q_k$ are real parameters which have to be fixed.
As in the case of the supercharges (\ref{plyushchay:Qf}),
the requirement of conservation of (\ref{plyushchay:QP})
results in the Hamiltonian (\ref{plyushchay:Hq1}) but with
the exponential superpotential
\cite{plyushchay:klishevich&plyushchay01a}:
\begin{align}\label{plyushchay:WExp}
 W(x)&=w_+e^{\omega x}+w_-e^{-\omega x}+w_0,&
 \omega^2&=-\frac{24q}{n\left(n^2-1\right)},&&&
\end{align}
where all the parameters $w_{\pm,0}$ are real, while
the parameter $\omega$ is real or pure imaginary
depending on the sign of the real parameter $q$,
and for the sake of simplicity
we put $\hbar=1$. In the limit $\omega\to 0$ this
superpotential is reduced to the quadratic form
(\ref{plyushchay:Hq1})
via the appropriate rescaling of the parameters
$w_{\pm,0}$.

The family of
$n$-supersymmetric systems given by the superpotential
(\ref{plyushchay:WExp}) is tightly related to the so called
quasi-exactly solvable problems
\cite{plyushchay:turbiner88,plyushchay:shifman,
plyushchay:ushveridze,
plyushchay:finkel&gonzalez&kamran&olver&rodriguez,
plyushchay:bender&dunne}.
Indeed, the both of the
Hamiltonians constituting the supersymmetric Hamiltonian of
the
form (\ref{plyushchay:Hq1}) with the exponential
superpotential belong
to the $sl(2,\mathbb R)$ scheme
of one-dimensional QES systems
\cite{plyushchay:turbiner88,plyushchay:shifman,
plyushchay:ushveridze}.
Besides, the QES family
given by the superpotential (\ref{plyushchay:WExp}) is
related  to
the exactly solvable Morse potential for some choice of the
parameters
\cite{plyushchay:klishevich&plyushchay01a}.

The $n=2$ non-holomorphic supersymmetry corresponding
to Eqs. (\ref{plyushchay:CalogeroQ}),
(\ref{plyushchay:CalogeroH})
occupies an especial position. Like the linear
supersymmetry, it admits the anomaly-free  quantum
formulation in terms of an arbitrary superpotential. Indeed,
the quantization of the supersymmetric system
(\ref{plyushchay:CalogeroH}) leads to
\cite{plyushchay:klishevich&plyushchay01a}
\begin{align}
 H&=\frac 12\left[-\hbar^2\frac{d^2}{dx^2}+W^2
 -\frac c{W^2}+v+2\hbar W'\sigma_3+\Delta(W)\right],
 \label{plyushchay:HD}\\
 Q^+&=(Q^-)^\dagger=
 \frac 12\left[\left(\hbar\frac d{dx}+W\right)^2
 +\frac c{W^2}-\Delta(W)\right]\theta^+,
 \label{plyushchay:QD}\\
 \Delta&=\frac{\hbar^2}{4W^2}\left(2W''W-W'{}^2\right).
 \label{plyushchay:WD}
\end{align}
Looking at the quantum Hamiltonian (\ref{plyushchay:HD})
and supercharges (\ref{plyushchay:QD}), we see that the
presence of the
quadratic in $\hbar^2$ term (\ref{plyushchay:WD}) in the
operators $H$
and $Q^+$ is crucial for preserving the supersymmetry
at the quantum level. Therefore, one can say that
the quantum correction (\ref{plyushchay:WD}) cures the
problem of the quantum anomaly since without it the
operators $Q^\pm$  would not be the integrals of motion.
The supercharges (\ref{plyushchay:QD}) satisfy the relation
$
\{Q^+,Q^-\}=(H-v)^2+c,
$
and the structure of the lowest bounded states in the cases
$c>0$,
$c<0$ and $c=0$ for $v=0$ is reflected  in the table and on
the figure
(for the details see Ref.
\cite{plyushchay:klishevich&plyushchay01a}).

\begin{figure}[h]
\begin{center}
\begin{picture}(280,100)
\put(-2,90){E}
\put(0,10){\line(1,0){50}}
\put(10,0){\vector(0,1){100}}
\put(25,-5){a)}
\put(20,30){\circle* 4}
\put(30,30){\circle* 4}

\put(20,50){\circle* 4}
\put(30,50){\circle* 4}

\put(20,70){\circle* 4}
\put(30,70){\circle* 4}

\put(20,90){\circle* 4}
\put(30,90){\circle* 4}

\put(105,-5){b)}
\put(80,10){\line(1,0){50}}
\put(90,0){\vector(0,1){100}}
\put(100,10){\circle 4}

\put(100,30){\circle* 4}
\put(110,30){\circle* 4}

\put(100,50){\circle* 4}
\put(110,50){\circle* 4}

\put(100,70){\circle* 4}
\put(110,70){\circle* 4}

\put(100,90){\circle* 4}
\put(110,90){\circle* 4}

\put(185,-5){c)}
\put(160,10){\line(1,0){50}}
\put(170,0){\vector(0,1){100}}
\put(180,10){\circle 4}
\put(190,10){\circle 4}

\put(180,30){\circle* 4}
\put(190,30){\circle* 4}

\put(180,50){\circle* 4}
\put(190,50){\circle* 4}

\put(180,70){\circle* 4}
\put(190,70){\circle* 4}

\put(180,90){\circle* 4}
\put(190,90){\circle* 4}

\put(265,-5){d)}
\put(240,10){\line(1,0){50}}
\put(250,0){\vector(0,1){100}}
\put(260,10){\circle 4}
\put(260,30){\circle 4}

\put(260,50){\circle* 4}
\put(270,50){\circle* 4}

\put(260,70){\circle* 4}
\put(270,70){\circle* 4}

\put(260,90){\circle* 4}
\put(270,90){\circle* 4}
\end{picture}
\vskip 5mm
\parbox{13cm}{
Figure. The four types of the spectra for the $n=2$
supersymmetry for bounded \\
\hspace*{1.2cm} states.}
\end{center}
\end{figure}

\begin{table}[h]
\begin{center}
\parbox{13cm}{Table. The structure of the lowest states for
the
$n=2$ supersymmetry.}
\vskip 2mm
\begin{tabular}{|ll|c|c|c|}
\hline
&&$c>0$&$c=0$&$c<0$
\\\hline
{\bf a)}&
Completely broken phase, there are no singlet
states &+&+&+
\\\hline
{\bf b)}& One singlet state in either bosonic or fermionic
sector
&&+&+
\\\hline
{\bf c)}&
\parbox{9cm}{\baselineskip=0.5mm
Two singlet states with $E=0$, one is in
fermionic sector, another is in bosonic sector}
&&+&+\\\hline
{\bf d)}&
Two singlet states in one of two sectors&&&+
\\\hline
\end{tabular}
\end{center}
\end{table}

{}From this structure one can  see, in particular, that  the
quantum theory ``remembers'' its classical origin:
the case $c>0$ corresponding classically to the holomorphic
$n=0$ supersymmetry gives the systems
in the completely broken phase
for any superpotential providing the existence of bounded
states.

In conclusion of the discussion of the nonlinear
supersymmetry
for the 1D quantum systems,
we note that for the first time the close
relationship between the nonlinear
supersymmetry and QES systems was observed in
Ref.~\cite{plyushchay:klishevich&plyushchay01a}.
Recently, it was demonstrated
\cite{plyushchay:aoyama&nakayama&sato&tanaka} that
the so called type A $\cal N$-fold supersymmetry
\cite{plyushchay:aoyama&sato&tanaka}
being  a generalization of the one-dimensional holomorphic
supersymmetry  is, in essence, equivalent to
the one-dimensional QES systems associated with
the $sl(2,\mathbb R)$ algebra.

\section*{Nonlinear supersymmetry on plane in magnetic
field}
The nonlinear holomorphic supersymmetry we have discussed
has a universal nature due to the  algebraic construction
underlying it and revealed in Ref.
\cite{plyushchay:klishevich&plyushchay01b}.
This universality allows us, in particular, to generalize
the above
analysis to the case of the two-dimensional systems.

The classical Hamiltonian of a charged spin-$1/2$ particle
($-e=m=1$) with gyromagnetic ratio $g$ moving on a plane and
subjected to a magnetic field $B({\bf x})$ is given by
\begin{equation}\label{plyushchay:Hcl}
 H=\frac 12\boldsymbol{\mathcal P}^2
 + gB({\bf x})\theta^+\theta^-,
\end{equation}
where $\boldsymbol{\mathcal P}={\bf p}+{\bf A}({\bf x})$,
${\bf A}({\bf x})$ is a 2D gauge potential,
$B({\bf x})=\partial_1A_2-\partial_2A_1$.
The variables $x_i$, $p_i$, $i=1,2,$
and complex Grassman variables $\theta^\pm$,
$(\theta^+)^*=\theta^-$,
are canonically conjugate with respect to the
Poisson-Dirac  brackets,
$\{x_i,p_j\}_*=\delta_{ij}$,
$\{\theta^-,\theta^+\}_*=-i$.
For even values of the gyromagnetic ratio $g=2n$,
$n\in\mathbb N$,
the system (\ref{plyushchay:Hcl}) is endowed with the
nonlinear $n$-supersymmetry. In this case
the Hamiltonian (\ref{plyushchay:Hcl})
takes the form
\begin{align}\label{plyushchay:Hz}
 H_n&=\frac 12Z^+Z^- + \frac i2n
 \left\{Z^-,Z^+\right\}_*
 \theta^+\theta^-,&
 Z^\pm&=\mathcal P_2\mp i\mathcal P_1,
\end{align}
which admits the existence of the odd integrals of motion
\begin{equation}\label{plyushchay:Qcl}
 Q^\pm=2^{-\frac n2}(Z^\mp){}^n\theta^\pm
\end{equation}
generating the nonlinear $n$-superalgebra
(\ref{plyushchay:aclas}).
The $n$-superalgebra does not depend on the explicit form
of the even complex conjugate variables $Z^\pm$. Therefore,
in principle, $Z^\pm$ can be arbitrary functions of the
bosonic dynamical variables of the system.

The nilpotent quantity $N=\theta^+\theta^-$
is, as in the 1D case, the even integral of motion.
When the gauge potential ${\bf A}({\bf x})$
is a 2D vector, the system (\ref{plyushchay:Hz}) possesses
the additional even integral of motion
$L=\varepsilon_{ij}x_ip_j$.
The integrals $N$ and $L$ generate the $U(1)$ rotations of
the odd, $\theta^\pm$, and even, $Z^\pm$, variables,
respectively. Their linear combination $J=L+nN$
is in involution with the supercharges,
$\{J,Q^\pm\}_*$=0,
and plays the role of the central charge of the classical
$n$-superalgebra. As we shall see, at the quantum level
the form of the nonlinear $n$-superalgebra
(\ref{plyushchay:aclas})
is modified generically by the appearance of the
nontrivial central charge in the anticommutator
of the supercharges.

A spin-1/2 particle moving on a plane in a constant
magnetic field represents the simplest case of a quantum 2D
system admitting the nonlinear supersymmetry.
Such a system corresponds to the $n$-supersymmetric quantum
oscillator
\cite{plyushchay:klishevich&plyushchay01b}. As in the
case of the one-dimensional theory, the attempt to
generalize the $n$-supersymmetry of the system to the case
of the magnetic field of general form faces the problem of
quantum anomaly. The generalization is nevertheless possible
for the magnetic field of special form
\cite{plyushchay:klishevich&plyushchay01b}.

To analyse the nonlinear $n$-supersymmetry for
arbitrary $n\in \mathbb N$,
it is convenient to introduce the complex
oscillator-like operators
\begin{align}\label{plyushchay:Z}
 Z&=\partial+W(z,\bar z),&
 \bar Z&={}-\bar\partial+\bar W(z,\bar z),&&&&&
\end{align}
where the complex superpotential is defined by
$\mathop{\sf Re}W=A_2({\bf x})$,
$\mathop{\sf Im}W=A_1({\bf x})$,
and the notations $z=\frac 12(x_1+ix_2)$,
$\bar z=\frac 12(x_1-ix_2)$, $\partial=\partial_z$,
$\bar\partial=\partial_{\bar z}$ are introduced.

The magnetic field is
defined by the relation $[Z,\bar Z]=2B(z,\bar z)$.
The $n$-supersymmetric Hamiltonian has the form
\begin{equation}\label{plyushchay:Halg}
 H_n=\frac 14\left\{\bar Z,Z\right\}+
 \frac n4\left[Z,\bar Z\right]\sigma_3.
\end{equation}
For $n=1$ we reproduce the usual supersymmetric Hamiltonian.
Unlike the linear supersymmetry, the nonlinear holomorphic
supersymmetry exists only when the operators
(\ref{plyushchay:Z}) obey
the relations
\begin{align}\label{plyushchay:int}
 \left[Z,\left[Z,\left[Z,\bar Z\right]\right]\right]&=
 \omega^2\left[Z,\bar Z\right],&
 \left[\bar Z,\left[\bar Z,\left[Z,\bar Z\right]\right]
 \right]&= \bar\omega^2\left[Z,\bar Z\right].&
\end{align}
Here $\omega\in\mathbb C$ and $\bar\omega=\omega^*$.
Using Eq. (\ref{plyushchay:int}), one can prove
algebraically
by the mathematical induction that for the system
(\ref{plyushchay:Halg}) the odd operators defined by the
recurrent
relations
\begin{align}\label{plyushchay:Qalge}
 Q^+_{n+2}&=\frac 12\left(Z^2-\left(\tfrac{n+1}2
 \right)^2
 \omega^2\right)Q^+_n,&
 Q_0^+&=\theta^+,& Q_1^+&=2^{-\frac 12}Z
 \theta^+,&
\end{align}
are the integrals of motions, i.e. they are supercharges.
One
can make sure
\cite{plyushchay:klishevich&plyushchay01a} that
in the 1D case these
operators generate the
nonlinear supersymmetry with the polynomial
superalgebra~(\ref{plyushchay:NSUSY}).

In the representation (\ref{plyushchay:Z}) the conditions
(\ref{plyushchay:int})
acquire the form of the differential equations for
magnetic field:
\begin{align}\label{plyushchay:eqBz}
 \left(\partial^2-\omega^2\right)B(z,\bar z)&=0,&
 \left(\bar\partial^2-\bar\omega^2\right)B(z,\bar z)&=0.&
\end{align}
The general solution to these equations is
\begin{equation}\label{plyushchay:Bz}
 B(z,\bar z)=w_+e^{\omega z+\bar\omega\bar z}+
 w_-e^{-\left(\omega z+\bar\omega\bar z\right)}+
 we^{\omega z-\bar\omega\bar z}+
 \bar we^{-\left(\omega z-\bar\omega\bar z\right)},
\end{equation}
where $w_\pm\in\mathbb R$, $w\in\mathbb C$,
$\bar w=w^*$.
On the other hand, for $\omega=0$ the
solution to Eq. (\ref{plyushchay:eqBz}) is the polynomial,
\begin{equation}\label{plyushchay:Bp}
 B({\bf x}) =
 c\left((x_1-x_{10})^2+(x_2-x_{20})^2\right)+c_0,
\end{equation}
with $c$, $c_0$, $x_{10}$, $x_{20}$
being some real constants. Though the latter solution can be
obtained formally from (\ref{plyushchay:Bz})
in the limit $\omega\to 0$ by rescaling appropriately the
parameters $w_\pm$, $w$,
the corresponding limit procedure is singular
and the cases (\ref{plyushchay:Bz}) and
(\ref{plyushchay:Bp})
have to be treated separately.

Since the conservation of the supercharges is proved
algebraically, the operators $Z$, $\bar Z$ can have any
nature (the action of $Z$, $\bar Z$ is supposed to be
associative). For example, they can have a matrix structure.
With this observation the nonlinear supersymmetry can be
applied to the case of matrix Hamiltonians
\cite{plyushchay:turbiner94,plyushchay:brihaye&kosinski,
plyushchay:finkel&gonzalez&rodriguez}.

Thus, the introduction of the operators $Z$, $\bar Z$
allows us to reduce the two-dimensional
holomorphic $n$-supersymmetry to the pure algebraic
construction. It is worth noting that in the literature the
algebraic relations (\ref{plyushchay:int}) are known as
Dolan-Grady
relations. The relations of such a form appeared for the
first time in the context of integrable models
\cite{plyushchay:dolan&grady}.

The essential difference of the $n$-supersymmetric 2D system
(\ref{plyushchay:Halg}) from the corresponding 1D
supersymmetric system
is the appearance of the central charge
\begin{equation}\label{plyushchay:Jz}
 J_n=-\frac 14\left(\omega^2\bar Z^2
 +\bar\omega^2Z^2\right)
 +\partial B\bar Z
 +\bar\partial BZ - B^2
 +\frac n2\bar\partial\partial B\sigma_3,
\end{equation}
$[H_n,J_n]=[Q_n^\pm,J_n]=0$.
The anticommutator of the
supercharges contains it for any $n>1$.
For example, the $n=2$ nonlinear superalgebra is
\begin{align}
 \left\{Q_2^-,Q_2^+\right\}&=H_2^2+\frac 14J_2 +
 \frac{\left|\omega\right|^4}{64}.
\end{align}
The systems (\ref{plyushchay:Halg}) with the magnetic field
(\ref{plyushchay:Bz})
of the pure hyperbolic ($w=0$) or pure trigonometric
($w_\pm=0$) form can be reduced to the one-dimensional
problems with the nonlinear holomorphic
supersymmetry \cite{plyushchay:klishevich&plyushchay01b}.

Let us turn now to the polynomial magnetic field
(\ref{plyushchay:Bp}).
One can see that this case reveals a nontrivial relation of
the holomorphic $n$-supersymmetry of the 2D system to the
non-holomorphic 1D $\cal N$-fold supersymmetry of Aoyama et
al
\cite{plyushchay:aoyama&sato&tanaka}.

In the system (\ref{plyushchay:Halg}) with the polynomial
magnetic field (\ref{plyushchay:Bp}) the central charge has
the form
\begin{equation}\label{plyushchay:Jp}
 J_n=\frac 1{4c}\left(\partial B(z,\bar z)\bar Z
 +\bar\partial B(z,\bar z)Z-B^2(z,\bar z)
 +\frac n2\bar\partial\partial B(z,\bar z)
 \sigma_3\right).
\end{equation}
It can be obtained from the operator (\ref{plyushchay:Jz})
in the limit
$\omega\to 0$ via the same rescaling of the parameters of
the exponential magnetic field which transforms
(\ref{plyushchay:Bz})
into (\ref{plyushchay:Bp}). The essential feature of this
integral is
its linearity in derivatives.

The polynomial magnetic field (\ref{plyushchay:Bp}) is
invariant under rotations
about the point $(x_{10},x_{20})$.
Therefore, one can expect that the operator
(\ref{plyushchay:Jp}) should be related to
a generator of the axial symmetry.
To use the benefit of this symmetry,
one can  pass over to the polar coordinate system
with the origin  at the point $(x_{10},x_{20})$.
Then the magnetic field is radial, $B(r)=cr^2+c_0$.
The supercharges have the simple structure:
$Q_n^+=2^{-\frac n2}Z^n\theta^+=(Q_n^-)^\dag$.
As in the case $\omega\ne 0$, the anticommutator of the
supercharges is a polynomial of the $n$-th degree in $H_n$,
$\{Q_n^-,Q_n^+\}=H_n^n+P(H_n,J_n)$, where $P(H_n,J_n)$
denotes a polynomial of the $(n-1)$-th degree. For example,
for $n=2$ one has
\[
 \{Q_2^-,Q_2^+\} = H_2^2 + cJ_2.
\]
For the radial magnetic field it is convenient to use the
gauge
\begin{equation}\label{plyushchay:Ap}
 A_\varphi=\frac 14cr^4+\frac 12c_0r^2,
 \qquad A_r=0.
\end{equation}
In this gauge the Hamiltonian (\ref{plyushchay:Halg}) reads
\begin{equation}\label{plyushchay:hb1}
 H_n={}-\frac 12\left(\partial_r^2+r^{-1}\partial_r
 -r^{-2}\left(A_\varphi^2(r)-2iA_\varphi(r)
 \partial_\varphi -\partial_\varphi^2\right)\right)
 +\frac n2B(r)\sigma_3,
\end{equation}
while the central charge (\ref{plyushchay:Jp}) takes the
form
$J_n={}-i\partial_\varphi-\frac{c_0^2}{4c}+\frac n2
\sigma_3$. Thus, the integral $J_n$ is associated with the
axial symmetry of the system under consideration.
The simultaneous eigenstates of the operators $H_n$ and
$J_n$ have the structure
\begin{equation}\label{plyushchay:ps2}
 \Psi_m(r,\varphi)=
 \begin{pmatrix}
 e^{i(m-n)\varphi}\chi_m(r)\\[2mm]e^{im\varphi}\psi_m(r)
 \end{pmatrix}.
\end{equation}

Since the angular variable $\varphi$ is cyclic, the 2D
Hamiltonian (\ref{plyushchay:hb1}) can be reduced to the 1D
Hamiltonian. The kinetic term of the Hamiltonian
(\ref{plyushchay:hb1})
is Hermitian with respect to the measure $d\mu=rdrd\varphi$.
In order to obtain a one-dimensional system with the usual
scalar product defined by the measure $d\mu=dr$, one has to
perform the similarity transformation $H_n\to UH_nU^{-1}$,
$\Psi\to U\Psi$ with $U=\sqrt r$. Since the system obtained
after such a transformation is originated from the
two-dimensional system, one should keep in mind that
the variable $r$ belongs to the half-line, $r\in[0,\infty)$.
After the transformation, the reduced one-dimensional
Hamiltonian acting on the lower (bose) component of the
state (\ref{plyushchay:ps2}) reads as
\begin{equation}\label{plyushchay:x6}
 {\cal H}_n^{(2)}={}-\frac 12\frac{d^2}{dr^2}
 + \frac{c^2}{32}r^6 + \frac{c_0c}8r^4
 + \frac 18\left(c_0^2-2c(2n - m)\right)r^2
 + \frac{m^2-\frac 14}{2r^2}
 -\frac 12(n-m)c_0.
\end{equation}
This Hamiltonian gives the well-known family
of the quasi-exactly solvable systems
\cite{plyushchay:turbiner88,plyushchay:shifman,
plyushchay:turbiner94,plyushchay:bender&dunne}.
The superpartner ${\cal H}_n^{(1)}$ acting on the upper
(fermi) component of the state (\ref{plyushchay:ps2}) can be
obtained
from ${\cal H}_n^{(2)}$ by the substitution $n\to-n$,
$m\to m-n$.

The reduced supercharges have the form
\begin{align*}
 {\cal Q}_n^+&=2^{-\frac n2}{\cal Z}_n\theta^+
 =({\cal Q}_n^-)^\dag,&
 {\cal Z}_n&=\left(A-\tfrac{n-1}r\right)
 \left(A-\tfrac{n-2}r\right)\ldots A,&&
\end{align*}
where $A=\frac d{dr}+W(r)$ and the superpotential is
\[
 W(r)=\frac 14cr^3+\frac 12c_0r+\frac{m-\frac 12}r.
\]
The operators ${\cal Q}_n^\pm$, ${\cal H}_n^{(i)}$,
$i=1,2$, generate
the non-holomorphic type A ${\cal N}$-fold supersymmetry
discussed in
\cite{plyushchay:aoyama&sato&tanaka}. The supersymmetry is
exact for
$c>0$ ($c<0$) and corresponding zero modes of the
supercharge
${\cal Q}_n^+$ (${\cal Q}_n^-$)
can be found.
The relation of the ${\cal N}$-fold supersymmetry with the
cubic superpotential to the family of QES system
(\ref{plyushchay:x6})
with the sextic  potential was also noted in Ref.
\cite{plyushchay:dorey&dunning&tateo}.

\section*{Resume}

To conclude, let us summarize the main  results of our
consideration of the nonlinear supersymmetry.

\begin{itemize}
\item
Generalized statistics and supersymmetry
are intimately related.

\item
 Linear supersymmetry at the classical level is a particular
 case of a classical
 supersymmetry characterized by the Poisson algebra being
 nonlinear in Hamiltonian.

\item
 Any classical 1D supersymmetric system is symplectomorphic
 to the supersymmetric system of the canonical form
 characterized by the
 holomorphic supercharges. There are three different classes
 of the classical canonical supersymmetric systems defined
 by the behaviour of the superpotential.

\item
 The anomaly-free quantization of the classical 1D
 holomorphic
 $n$-supersymmetry is possible for the quadratic and
 exponential superpotentials.

 \item
 The nonlinear supersymmetry is closely related to the
 quasy-exactly solvable systems.

\item
 The $n=2$ supersymmetric Calogero-like systems
 (\ref{plyushchay:CalogeroH}) admit the
 anomaly-free quantization for any superpotential;
 the specific quantum term ($\sim\hbar^2$)
 ``cures'' the quantum anomaly problem.

\item
 The anomaly-free quantization of the classical 2D
 holomorphic
 $n$-supersymmetry fixes the form of the magnetic field to
 be the quadratic or exponential one.

\item
 Realization of the holomorphic $n$-supersymmetry in 2D
 systems leads to the appearance of the central charge
 entering
 nontrivially into the superalgebra.

\item
The holomorphic nonlinear supersymmetry
can be related to other
known forms  of nonlinear supersymmetry
via the dimensional reduction procedure.

\item
 There is the universal algebraic foundation
 associated with the Dolan-Grady relations
 which underlies the holomorphic $n$-supersymmetry.

\end{itemize}

The universal algebraic structure underlying the
holomorphic nonlinear supersymmetry
opens the possibility  to apply the latter
for investigation of the wide class of the quantum
mechanical systems including the models described by the
matrix
Hamiltonians, the models on the non-commutative space,
and integrable models
\cite{plyushchay:klishevich&plyushchay01c}.

\subsection*{Acknowledgements}
M.P. thanks the organizers for hospitality
and A. Zhedanov for useful discussions.
The work was supported by the grants 1010073 and 3000006
from FONDECYT (Chile) and by DICYT (USACH).

\end{document}